# Consumer belonging behaviour: Qualitative testing of a theoretical framework and proposal of an alternative model

**Lizardo Vargas-Bianchi**
Graduate School, Universidad de Lima
lvargas@ulima.edu.pe

**Abstract**
Much research has been conducted on how consumption is related to human relations, for example, consumer communities organized around specific brands, or the way people use products to define their own identity and transmit a desired image. However, only a scarcity of research has examined the consumption behaviour when the fundamental intention is to leverage group belonging. The literature comprises a single theoretical framework that describes this behaviour, a nascent proposition that has not been tested. This study examines the transferability of that theoretical framework in a different context than the one used for its proposal and its extent on the phenomenon of consuming to leverage belonging. A qualitative deductive case study and a pattern matching analysis technique were employed, followed by a structural coding analysis of interview data. The findings revealed the model is transferable, however its conceptual extent on the phenomenon it addresses faces limitations. These findings allow the proposal of an alternative framework, the Belonging-Oriented Consumption Model. This model provides a theoretical basis for future research on consumer belonging behaviour.

**Introduction**
As social beings, we live immersed in relationships with one another that define our identity and behaviour within the cultural framework in which we live. Consumption is part of that social reality, which is noted in peers' influence on purchasing decisions, as well as the assessment of products and brands based on the opinions of acquaintances (Schulz, 2015). Researchers argue that consumption is related to the desire to belong, since purchasing behaviours do not occur in isolation from social ties and affiliation with family, community, or other groups (Arvidsson & Caliandro, 2016; Bettman & Escalas, 2005; McAlexander, Schouten & Koening, 2002; Muniz & O'Guinn, 2001). Studies show that consumers relate to each other using of specific brands and enjoy satisfaction through shared emotions (Ruane & Wallace, 2015). For example, brand communities have revealed how individuals relate in situations where a brand is the foundation of their encounter (Goulding, Shankar, & Canniford, 2013; Grzeskowiak & Sirgy, 2007). Some authors have researched extensively how individuals use products to reflect their own identity. Belk's work on the extended self through symbolic incorporation of object meanings is one of the most influential studies (Belk, 1988, 2017). He argues the individual relates to some brands as if they reflected himself, as a means for self-representation and a collective conception of self. While this connects to the fact that people present and share their identity with others, it does not consider the steps that people follow when they intentionally select and use possessions to support their affiliation to a group. The literature on this topic coincides that the search for satisfying belongingness gives rise to a series of behaviours that are influenced by other people who are members of a group to which they belong, or wish to belong, These conducts can, in opportunities, influence purchasing choices. Despite related studies, not much research has modelled behaviour by describing the stages entailed when people use products or brands to consciously leverage group belonging. No accepted model is found in academic literature that describes this consumption action and allows for future analysis (Arias, 2016; Arias & Otnes, 2016, 2017; Mead et al., 2011). Arias & Otnes (2016, 2017) proposed an initial theoretical framework entitled Belonging Process (BP) that explains the process that individuals follow when they consume to meet their need of belonging. The study aims to test the transferability of this nascent model, besides examining its scope to provide an understanding of consumption related to group belonging, and furthermore to propose an alternative model that addresses those aspects of the phenomenon that are not yet covered. Consequently, the study comprised two research questions:

> RQ1: Does the BP theoretical framework describe and explain consumption behaviours that aim to leverage individuals' group belonging in a cultural and time context that differs from the original study?

> RQ2: Does the conceptual scope of the BP theoretical framework comprehensively address the phenomenon of consumption behaviour that individuals engage in to promote their group affiliation?

This study contributes to the understanding of how individuals deliberately take advantage of consumer activities to benefit their group belonging, as well as purchasing decisions which are not based on the functional characteristics of a product, or the value propositions expressed by each brand, but on the social value that the individual gives to goods available in the market. The findings of this study contribute to the growing body of knowledge about person-object-person relationships (Belk, 1988; McCracken, 1986) and belonging-related consumer behaviour.

**Literature review**
Since the development of Maslow's motivational scheme and the contributions of Bowlby's attachment theory, belonging has been present in human behaviour studies as a condition to ensure well-being (Grzeskowiak & Sirgy, 2007) and as an influence on interpersonal relations (Nichols & Webster, 2013). Belonging studies describe the concept as the human need to experience belongingness and acceptance in a group and by its members; it may be regarded as a reciprocal dynamic because group fellows expect the individual to reciprocate the attention (Mellor et al., 2008; Pickett, Gardner, & Knowles, 2004). Scholars have argued that the desire to belong influences the purchase decision of a variety of product categories including clothing, travel destinations and consumer electronics. These products assist people in taking care of their social relationships (Griskevicius, & Kenrick, 2013).

In their extensive literature review, Baumeister and Leary (1995) proposed a hypothesis about the need to belong (NTB) that has been a central influence in studies related to this subject. They noted that belonging is a fundamental human need, which explains much of social behaviour. The authors further stated that people have a strong innate inclination to form and maintain a minimum amount of stable and meaningful interpersonal relationships. They added that when NTB is not met, individuals experience a variety of negative emotional and cognitive effects. Baumeister and Leary noted that two conditions are required for this desire to be met: interactions with other people must be frequent and affectionately pleasant, and these interactions must occur in a lasting context and be of emotional concern and attention to others as the individuals must experience or believe that their peers are concerned with their well-being. Related research agrees that NTB is associated with people's sensitivity to the perception of signals that proof acceptance (or rejection) by members of a group. Some authors relate the desire of people seeking for group affiliation to the fact that they are attentive to any information that reveals something about their peers (Leary, et al., 2013; Pickett et al., 2004). Sheth, Newman, & Gross (1991) state that the individuals' purchase decision process includes a social variable. They call it social value, meaning the perceived utility of the product's association with specific groups. Brands or products can serve as guides for signalling approval, and thus can perform as instruments to assist relationships with others, and to enable affiliation (Berger & Heath, 2007; Hammerl et al., 2016; Ratner, Kahn, & Kahneman, 1999). Baumeister and Leary also described different dimensions with which this need may be related. Although they did not mention that purchase actions are related to belonging, they linked NTB to diverse behaviours that derive from this motivation. For example, recent studies have argued that the desire to belong manifests in people's behaviours on social media such as Facebook and Instagram when they share photographs and/or messages aimed at favouring their connection with groups of their interest (Nadkarni & Hofmann, 2012; Seidman, 2013).

Similarly, in group conformity, individuals purchase products to adhere to group norms and attempt to maintain homogeneity with the tastes of group members by their purchases (Beran, Kaba, Caird, & McLaughlin, 2014; Cialdini & Goldstein, 2004; Grotts & Johnson, 2013). Several studies on reference group influences have suggested a relation between group affiliation and the use of specific brands by its members (Bearden & Etzel, 1982; Bearden et al., 1989; Bettman & Escalas, 2005; Burnkrant & Cousineau, 1975; Childers & Rao, 1992). Authors establish a link between belongingness need and purchase actions in group conformity (Algesheimer, Dholakia, & Herrmann, 2005). This phenomenon is observed when individuals acquire products to adhere to group norms, or when their purchasing action is likely to be consistent with the tastes or opinions of group members, for example in the choice of a particular product model or brand (Griskevicius, Goldstein, et al. 2006; Park & Feinberg, 2010).

White and Argo (2009) argue that consumers avoid products that signal their membership of groups to which they associate with a negative perception. Various works that have analysed the influence of reference groups point out the relationship between the affiliation to a group and the use of trademarks by its members (Bearden & Etzel, 1982; Bearden et al., 1989; Burnkrant & Cousineau, 1975; Childers & Rao, 1992). This phenomenon is also noted, for example, when a person claims that he or she would buy a specific brand because friends or colleagues approve of it, or because he or she behaves according to the habits or rules followed by other members of the community (Algesheimer, Dholakia, & Herrmann, 2005). Veloutsou and Moutinho (2009) have stated that, because of the influence of the group, sometimes brand loyalty can be confused with group loyalty, when the individual's behaviour is primarily identified with that of the group.

Research has also described the effect between individuals within the so-called reference groups. A reference group is a collective, real or imaginary, that exerts a significant relevance in the evaluations, aspirations or conduct of the subject (Grzeskowiak & Sirgy, 2007; Park & Lessig, 1977). Within the framework of the reference group, consumers likewise find a source of information on ratings or attitudes from other people who share similar beliefs (Bettman & Escalas, 2005). The influence that the reference group exerts on individuals can include perceptions about a specific brand (Ruane & Wallace, 2015; Veloutsou & Moutinho, 2009), for example, when a product represents the beliefs or attitudes of the group members. It can happen that the consumer does not even like the product, and yet may wish to purchase it to gain acceptance from members of his or her reference group (Belk, 2017; Grotts & Johnson, 2013).

Belk (1988, 2017) argues that individuals incorporate symbolic meaning to possessions, so that certain objects or places become part of his personality. The objects allow their owner to elaborate a concept of "who I am", making them a means for personalization and self-representation. The researcher explains these relationships are not composed of only two elements (person-thing), but that three intervene: person-thing-person. This relationship extends to the meaning that people give to products as long as it allows them to link up with other subjects and find benefits in this relationship. Ahuvia (2015) points out that the formula 'person-thing-people' is a concise description of the phenomenon that links individuals and products and calls it "the first axiom of Belk". The author relates the extended self with group affiliation by arguing that one way to express it is through shared symbols of consumption. These symbols define what he calls the group self. Products such as clothing, cars, jewellery, or attendance at clubs or entertainment events express an individual sense of self, and also point out group identity. Belk also mentions collectively owned products as ways of demonstrating group affiliation, although he does not elaborate on the process people follow when seeking group belongingness.

The suggestion that consumption is often a person-thing-person relationship, or client-brand as proposed by Muniz & O'Guinn (2001), is witnessed in the so-called brand communities, where a specific brand functions as the axis that gives rise to interpersonal closeness and grounds the sense of belonging between individuals in that community (Algesheimer et al., 2005; McAlexander, Schouten, & Koenig, 2002; Muniz & O'Guinn, 2001). Brand communities are a set of social relationships that are structured around the use of a particular brand (Ahuvia, 2005; Berger & Heath, 2007; Bettman & Escalas, 2005; Goulding et al, 2013), between members who share a system of values, standards and representations (Cătălin & Andreea, 2014), and who have an identity based on that brand (McAlexander et al., 1981), or around shared ownership of a product (Ahuvia, 2005; Ruiz, 2005). Some authors refer to brand tribes rather than communities, although there does not seem to be any consensus on the differences between the two terms, and they are used interchangeably (Cova & Pace, 2006).

**Belonging Process theoretical framework**
Consumption and group affiliation have been a central question in the study of consumer culture and social processes. Research on brand communities, and on the use of possessions to constitute a person's identity, shed light on this phenomenon. However, they do not descend to describing the process that consumers follow when marketed goods are consciously used to promote belongingness. A few studies have identified this gap and analysed the proactive search for belonging according to which people instrumentalize consumption (Mead et al., 2011). Arias and Otnes (2016, 2017) conducted exploratory work to clarify how people take advantage of consumption to achieve favourable results related to their sense of belonging and proposed a theoretical framework to analyse this phenomenon. Their qualitative analysis, based on narratives from university students, formed the basis of the Belonging Process (BP) theoretical framework.

In their work, Arias & Otnes (2016, 2017) give special relevance to arguing that belonging is not a fixed state of practice, as postulated by studies on NTB (Baumeister & Leary, 1995), but that it is a procedural and fluid phenomenon. So much so they call their theoretical scheme the Belonging Process (BP). The authors affirm this process is composed of several constructs, which occur sequentially. Their writings seem to have a double purpose: to argue that belonging is a fluid reality, and to analyse consumer behaviour when their intention is to gain a sense of belonging. The present study does not intend to examine whether the sense of belonging is in all cases fixed or procedural. A description of the six constructs the BP comprises follows.

Cultural context: Arias and Otnes (2016, 2017) noted that it includes the broadest drives that shape the BP framework. It includes both sociological and cultural forces such as social norms or built expectations among the members of that context. Its role in the belonging process is to encompass its influences on each construct and the way they are related entirely.

Belonging targets: Arias and Otnes (2016, 2017) argued that belonging targets comprise the social entities to which the individual seeks to belong, to be a member of. They may exist on three levels: micro, meso and macro. These targets place belonging within individuals' context in which their behaviour occurs. Belonging targets influence the identification of the belonging conduits and provide belonging cues, so consumers can gauge their status concerning group belonging. The effect can be direct, such as consumption suggestions received or heard from peers of the target group or indirect, for example, when the subject notices the consumption patterns of the target group members.

Belonging conduits: These are the resources available on the market that consumers employ to leverage their belonging status of a group. They include products, services, and brands. Their use depends on their symbolic or functional properties. Arias and Otnes (2016, 2017) explained that the way a product or experience is consumed may also serve as a conduit. As noted previously, BP proponents have proposed that belonging targets influence the selection that people make regarding conduits. Once the group to which individuals wish to belong is identified, they distinguish the consumption means that lead them to that goal. Identifying consumer products and/or behaviours may be directly influenced by group members.

Belonging cues: These include the cues or signals those consumers receive from the target group members that assist them to acknowledge their progress toward their belonging goal. Cues are internal when they originate from the person's own subjective perception and external when they originate from a third party. An example of the former is when consumers buy products they consider being exclusive or consumed by the elite. An example of the latter occurs when individuals receive compliments from group members because of something they are wearing or planning to buy.

Belonging obstacles: Arias and Otnes (2016, 2017) defined these as forces that prevent people from achieving their expectations of belonging to the target group. They argued that experiences of social exclusion originate not only from interpersonal relationships but are also because of personal and psychological factors that can hinder or even limit individuals' progress toward their goal of belonging.

Belonging results: These are also known as outcomes and include those experiences of feelings of belonging that the individuals achieve in relation to the belonging target. Outcomes may range from simple interactions with people that belong to the group to the establishment of stable relationships.

Arias and Otnes (2016, 2017) stated that their proposed framework is a preliminary theorization about consumption and its relationship with the desired sense of belonging. BP comprises the elements that are present in people's behaviour when proactively consuming goods with the primary aim of leveraging their belonging status. According to Arias and Otnes, this framework and its constructs afford academics a vocabulary with precise theoretical content that allows further research on belonging-related consumption.

**Methods**

A qualitative deductive method, which seeks to analyse a theory through testing and the review of its results, was employed to confirm or reject the BP theoretical framework (Creswell, 2007; Hyde, 2000; Løkke & Sørensen, 2014; Miller & Crabtree, 2005; Bitektine, 2008). Whether the theory is transferable to a context different from the original is ascertained (Yin, 2018). This methodological procedure requires the researcher to work on an existing theory, collect data to prove it, and subsequently, reflect on its confirmation (or rejection) grounded by the findings (Creswell, 2014). This study was exploratory. The latter approach is suitable for obtaining information from a theoretical field that is preliminary or in which there is partial knowledge (Bitektine, 2008; Creswell, 2007; Sinkovics, 2017).

**Study design.** A single case study design was adopted (Yin, 2018). The case study allows a phenomenon, referred to as a case, to be studied in depth and within its context, embracing the fact that the boundary between the phenomenon and the context is not clearly drawn (Creswell, 2007; Yin, 1981, 2006, 2018). In this study, the case was defined as the consumption behaviour consciously directed toward leverage group belonging. The case participants included 12 young females and males who were intentionally selected as they had consciously performed consumption actions to favour their belonging status. Arias and Otnes (2017) used a similar population in their original work in terms of age and dedication to their higher education responsibilities. Likewise, various studies have suggested that young people show greater sensitivity to group belonging and agreement with their peers (Hornsey & Jetten, 2004; Oishi, et al., 1999).

The case limits included: participants had to be aware of having made purchases and consumptions with the primary purpose of favouring their group belonging status during the previous year and they had to choose to purchase or use specific products or brands to assist their group belonging status. The time limit of the case was three months. The inclusion criteria for participant selection were: enrolment in a higher education institution; between the ages of 19 and 23 years; and classification as a medium, medium-high, or high socioeconomic level. An equal number of males and females were chosen, and their mean age was 22 years (median = 22; mode = 23; SD = 1.20).

All the participants had a Western background, belonged to a Latin America social and cultural frame of reference, and were immersed in a metropolitan urban lifestyle.

**Analysis technique.** A pattern matching technique (Trochim, 1985; Yin, 2018; Almutairi, Gardner, & McCarthy, 2014) following the procedure proposed by Pearse (2019) was employed in the analysis. In the pattern matching design, propositions articulated in the tested theory are linked with data obtained in the study by comparing a pattern of observed results with another pattern of the expected theory-derived values (Campbell, 1975; Hyde, 2000; Sinkovics, 2017; Bitektine, 2008). Hak and Dun (2010) referred to both extremes of information as the observed pattern and expected pattern because theories predict expected patterns, which are considered being hypotheses. Yin (2018) asserted that when the patterns revealed empirically coincide with the expected ones, the result helps to confirm these hypotheses and thus, the theory's validity is strengthened. A first cycle of descriptive coding analysis was conducted to have an in-depth understanding of the phenomenon under study within the limits of the tested theory (Saldana, 2013). Verbalization of consumption experiences directed to aid group belonging was established as the context unit to conduct the coding. A second cycle of analysis was carried out using a structural coding technique on the information collected from the participants. Structural coding applies the concepts that represent the research topic to the data set of a study (Saldana, 2013). The topics for this coding procedure were: (i) identification of the Target group; (ii) group affiliation behaviour; (iii) market resources used for group belonging; and (iv) recognition of belonging status. NVivo 13 was employed for both the pattern matching and coding analysis.

**Propositions**
To configure the expected pattern, seven BP theory-based propositions were formulated:

P1a–Belonging targets: The individuals whose consumption behaviours are intended primarily to leverage belonging declare that they identify and establish the groups to which they belong.
P1b - Belonging targets: The individuals identify and set the groups to which they seek to belong.
P1c - Belonging obstacles: The individuals identify factors that obstruct them from belonging to a current or future group.
P2a - Belonging conduits: The individuals identify products, services and brands or behaviours whose consumption will maintain their group belonging and projected affiliation to a belonging target.
P2b - Belonging conduits establishment: The individuals confirm that the belonging target directly or indirectly influences the establishment of belonging conduits.
P2c - Belonging cues type: The individuals confirm that the products, services and brands whose use leverage group belonging do so because of their functional or symbolic type.
P3 - Belonging cues: The individuals establish and identify the cues that show their belonging status relative to the belonging target.

**Codebook.** Under these propositions, the researcher developed a codebook to display the empirical data from the tested theory to allow the pattern matching analysis (Fereday & Muir-Cochrane, 2006; Pearse, 2019; Stuckey, 2015; Yukhymenko et al., 2014). The codebook is displayed in Table 1:

Table 1. Code book based on the BP theoretical framework of Arias & Otnes.

|  | Codes |
| --- | --- |
| P1a - Group identification | KNOWSGROUP - knows the groups it already belongs to |
| P1b - Belonging target | SEEKBTARGET - sets the groups it seeks to belong<br>MACRO - evidence that the target set is macro<br>MESO - evidence that the target set is meso<br>MICRO - evidence that the target set is micro |
| P1c - Belonging obstacles | IDOBSTAC - identifies factors that obstruct belonging to a target group |
| P2a - Conduits | IDPRODCOND - identifies a product that serves as a belonging conduit<br>IDCONDCOND - identifies a behaviour that serves as a belonging conduit<br>IDBRANDS - identifies a brand that serves as a belonging conduit |
| P2b - Selection of belonging conduits | TARGETINF - manifests (implicitly or explicitly) that target group influences the establishment of conduits |
| P2c - Character of the belonging conduits | FUNCT evidence that the belonging conduit is functional<br>SYMB –evidence that the belonging conduit is symbolic |
| P3 – Belonging cues | IDCUES – identifies the cues when they confirm group belonging |

**Data collection.** In-depth semi-structured interviews were conducted. At the researcher's request, teachers from three private higher education centres provided a list of students. They were asked via email to take part in the study. Those who responded in the affirmative received a digital questionnaire to evaluate whether they met the case limits and inclusion criteria. Those who complied were selected to be part of the group. Interviews were conducted telephonically. An inquiry guide (Table 2) based on the study propositions was used to guide the interviews, which lasted an average of 25 minutes. A professional service was used to transcribe the recorded interviews.

Table 2. In-depth enquiry guide and correspondence with the study propositions.

| Topics | Questions | Corresponding propositions |
|---|---|---|
| **Target groups belonging** | Have you identified the groups you belong to? | (P1a) |
| | Have you identified the groups you want to belong to? | (P1b) |
| **Conduit products/services/ brands for belonging** | Can you identify which products/services support your group belonging status? | (P2a) |
| | Can you identify which purchase/use behaviours support your group affiliation status? | (P2b) |
| | What information do you recognize that allows identification of those products/services/brands? | (P2a, P2b, P2c) |
| | Does anything or anybody define products that promote belonging status (to the target group)? | (P2b) |
| | Have you experienced or recognized obstacles in the quest for group belonging? | (P1c) |
| **Signs of belonging status** | How do you notice when a product/service/brand is favouring your group belonging? | (P3) |
| | Do you need someone to confirm the product/service/brand used is aiding group belonging? What happens if you don't get confirmation? | (P3) |

Three months after the interviews, all the participants were contacted by email and asked to complete a follow-up on-line questionnaire. The same topics in the interview guide (Table 2) were in the questionnaire. Participants answered each question on a five-point Likert scale, ranging from 1 (strongly disagree) to 5 (strongly agree). They also completed open-ended questions. The purpose of the follow-up questionnaire was a further source of information to assess the consistency of data got from the interviews. Six of the participants completed the questionnaire.

In relation to the ethical aspects of the study, the participants' anonymity and confidentiality were preserved by assigning a letter code to each informant. Participation was voluntary, and they all signed an informed consent form before their interviews. No participant declined to continue with the study or required to withdraw the information provided.

**Results**

Matches were revealed between all theoretical and empirical patterns in the BP theoretical framework, as well as some events of the consumption to leverage belonging behaviours that are not accounted for by the BP conceptions. The matching patterns and the unmet events are depicted in Table 3:

Table 3. Coincidence between predicted patterns, empirical patterns, and unmet events.

| Construct | Proposition | Theoretical pattern | Empirical pattern | Unmet events |
|---|---|---|---|---|
| Belonging Targets | P1a | The person manifests that he/she identifies and establishes the groups to which she/he belongs. | The person identifies and establishes the groups to which she/he belongs. | Identification of target group takes place in a non-systematic way; groups are defined based on roughly specified identities and a broad sense of belonging. |
| | P1b | The person states that she/he identifies and sets the groups to which she/he seeks to belong. | The person states that she/he identifies and sets the groups to which she/he seeks to belong. | Groups may be micro, meso, macro indistinctly and regardless of whether the participants recognized them that way. |
| | P1c | The person identifies factors that may be an obstacle to belonging to a target group. | The person identifies factors that may straightforwardly hinder obstacles to a target group. | There is no precise awareness of their occurrence; subjects do not seem to take much account of obstacles. |
| Belonging Conduits | P2a | The person identifies products/services/brands whose purchase or use he/she considers promotes the desired group belonging status. | The person identifies products/services/brands whose purchase or use he/she considers promotes the desired group belonging status. | Participants also identify other means available in the market that enhance the information and exhibition of the conduits. |
| | P2b | The person states that the belonging target directly or indirectly influences the establishment of the belonging conduits. | The person states that the belonging target directly or indirectly influences the establishment of the belonging conduits. | The belonging target does not appear as determining, since a level of flexibility or interpretation is perceived by the subject about available goods useful to relate to a group. |
| | P2c | The person shows that the products/services whose use favours belonging status do so because of a functional or symbolic nature. | The person shows that the products/services whose use favours belonging status do so because of a functional or symbolic nature. | This status does not seem to be an element to which they are aware of when choosing a good as a conduit. |
| Belonging Cues | P3 | The person establishes and identifies the cues that confirm his/her belonging status relative to the belonging target. | The person establishes and identifies the cues that confirm his/her belonging status relative to the belonging target. | The cues do not appear to be conditional, as the person can self-confirm belonginess when they consider that a product or consumer practice marks their affiliation in the group. |

**Category: Belonging targets**. The analysis revealed that the participants identified and established the groups to which they wished to belong (P1a). A participant thus expressed this: "*I feel a need to belong, for example, to work for one company in the financial sector (…) for a large company. I particularly feel that need to want to belong to that group, leave office hours and talk with people*" (Participant N, male). Another noted, "*Putting myself on the side of being fit, so* *now I am working a lot in the gym, exercising, changing to that lifestyle*" (Participant M, male). These targets had the properties of being macro, meso or micro groups, indistinctly and regardless of whether the participants recognize them that way.

Despite the match between the theoretical and empirical pattern, the coding analysis revealed that the participants do not engage in a schematic process to identify the groups with which they wish to be affiliated. Nor is there a detailed configuration of the identity of the group to which they wish to belong. On the contrary, they described the identities of the group they are interested in joining in a roughly or loosely defined manner, without precise delimitation. The BP model does not consider this way of proceeding among individuals.

**Category: Belonging obstacles**. The interviewees displayed some awareness that there are factors that can hinder participation in a target group (P1c). They shared experiences in which they identified problems in their efforts to belong. Participant R (male) argued, "*Before (…) he lifted less [weight in the gym] (…), however, he had all the motivation, he ate the same healthy foods, he trained, but he lifted less than that because he was just starting, and he was not part of the group even though he trained just like them, despite [doing] the same things*". One participant shared that if her close group of friends were consuming a product from a well-known coffee shop and she was not, she felt "*like we were all not going at the same pace*" (Participant A, female). It was perceived that the nature of the obstacles is broader than mere social exclusion, as stated by Arias and Otnes (2016, 2017), who explain that obstacles related to affiliation can have different origins and nature. The participants noted this when they said that an incorrect choice of conduits can prevent or damage their belonging effort. Participant J (male) shared an event where someone who tried to approach the group carried out consumer actions whose members perceived as factitious: "*[he] came into the group and tried to buy [a bottles of] whiskey, we really don't mind about that; he wanted to get into the group with that purchase and that was spotted*".

As with the Belonging targets, it can be noticed that Belonging obstacles are not always fully recognized by the interviewees. There is no precise awareness of their occurrence, subjects do not seem to take much account of them. Instead, people perceive that the real obstacle is not achieving the desired belongingness status, rather than attributing it to the material resources employed for their enterprise. The BP model disregards this unreflective way of considering, foreseeing and coping with possible obstacles.

**Category: Belonging conduits**. It was revealed that the participants found products and services that, when consumed, they considered favouring group belonging (P2a). One participant argued that people in a group "*usually use a lot of these [design of] black coats, I don't know exactly the brand (…), the people that most often use them are students of a particular university*" (Participant K, male). Another participant referred to products and belonging: "*Many times my friends also comment on a certain product, I also want to be part of the group and get to know how good [the product] is*" (Participant I, female). Arias and Otnes (2016, 2017) explained that the market offers a variety of possibilities for constituting conduits such as the use or constant mention of brand names, the use of similar products by group members and the display of logos and/or brand identifying signs.

The participants also gave examples of using conduits when they completed the questionnaire: "*Depending on which brands the majority [of people] use them to achieve similarity. With groups of [same] hobbies or [same] professions, it always favours having products of key brands that contribute to the work [they have]*" (Participant Y, female). A participant also referred to an additional feature, the geo-location of the purchased product when exhibited in a social media network, as information that can be used as a belonging conduit. He acknowledged sharing a consumer product used as a conduit when posting it on his Instagram account that included the purchase location in the story. Another participant stated: "*Even now you can put a location on it, so you not only show the cup [of Starbucks coffee] but where we are, where the brand is from*" (Participant E, female).

It is noteworthy that participants do not restrict themselves to instrumentalising products or brands as Belonging conduits, but any kind of market resource at their disposal to further their quest for affiliation. For instance, including Instagram's geo-location feature to include the location of the Starbucks shop in the same Instagram post where the product consumed is also shown, extends the definition of the Belonging conduits beyond its conceptualisation in BP.

The participants experienced consumer goods as products or as branded products as equally useful, provided they were means available in the market that could serve as a vehicle to meet their need to belong to the group of choice. One participant shared, "*I feel that if I buy something, for example, from Starbucks and my friends do too, I would feel that I am part of the group*" (Participant A, female). Participant R acknowledged, "*It is not the same to go on stage and play with Converse trainers that going with your Adidas or Reebok shoes, they do not go*". Sports apparel brands were often a recurring experience among brand mentions among the participants; for example, Participant K (male) related, "*As soon as you get to the court, the boys take out their rackets and during the halftime to go and see the racket's brand, what model it is (…) the same thing happens in soccer*".

*Participant C (female) illustrated this when she talked about how she used the brands: "Adidas, Nike, so (…) buying one of these makes me feel a bit more belonging, because my circle is full of people who practice sport".* Participant V (male) stated, "*Most of us belong to the design, advertising and creativity group, we all use Apple*". In this way, brands seem to become fit vessels to leverage the assembly of belonging relationships between people.

As the interviewees' experience shows, however, the Belonging targets do not emerge as a defining element when choosing Belonging conduits. As is noticeable in the preceding constructs, people address them in a flexible, non-conditional manner. The BP model states that the Belonging conduits will be configured in accordance with the Belonging targets. It does not seem to be the case that this happens in such a straightforward way.

Similarly, the participants stated the target group directly and indirectly influences which products or services may serve as conduits to belonging (P2b). A participant told he wanted to be part of a group, so he ate foods similar to those consumed by members of the group: "*Those who work in an office, usually opt for healthy food, so [to] integrate with them I also eat that healthy food and we all agree on eating healthy*" (Participant N, male). Another participant also shared her experience of the group's influence when she recounts: "My friends want to go to Starbucks and I have to go because I know they will take that picture with the cup" (Participant L, female).

Finally, P2c also matched the predicted theoretical pattern because the participants revealed that the products and/or services used as conduits for belonging do so because of their functional or symbolic nature.

As with Belonging obstacles, respondents were not concerned about the symbolic or functional status of the conduit when choosing it. While the BP model does not claim that this feature determines the choice or usage of a conduit, nor its relationship to the other constructs, it is pertinent to dimension its weight on belonging-oriented consumption behaviour.

**Category: Belonging cues**. In relation to the cues of signs (P3), the participants certified they could establish and identify those cues that show their belonging status concerning a specific group. One participant thus shared this: "*I use Adidas almost always and, somehow, it is like you wear Adidas shoes and they say 'Hey, you bought that pair', 'Yes. You know what? I saw them there, on a sale. Do you remember when we went to see them?'; so, it's like many times the conversation revolves around this product*" (Participant N, male). Participant X (female) said that she noticed when a specific product or consumption practice is leveraging its belonging status when "*[there is] a follow-up of posts [on Instagram], not only at once, but that it has a sequence, (…) That they ask me for recommendations maybe from places I attend to eat that kind of food*".

Based on the descriptive coding analysis performed, a phenomenon which was not specified within the BP propositions used in the pattern matching emerged among the participants. This phenomenon refers to the fact that some people are perceived as being closer to a group than others. They expressed it was possible for someone to have a more engrained or better-established belonging status within a group based on their consumption intensity of a product or brand. One participant thus explained this: "*In the gym group (…) there are some that enjoy buying powder supplements; (…) you could say those who buy them are like they have a 'plus' (…) Like they take it more seriously, they spend their money on supplements, a diet, those things*" (Participant R, male). A female participant noted, "*There are distinguishable groups, I have noticed that they are entirely Apple, iMac, iPhone, stuff like that*". She added that the identification of that group seems to be accurately portrayed–"*distinguishable groups*"–by the intensity of consumption of those branded products.

However, according to the information provided by the interviewees, the belonging cues originating from members of the group to which they wish to belong do not seem to be conditional. Participants state that the mere fact of consuming a product, or even having in mind to purchase it, can be a sufficient cue for confirmation as belonging to the group. Peer recognition does not seem to be a requirement, as subjects can self-confirm their affiliation to a group: if I already own this brand or product, I can consider myself part of the group of those who use it. In these cases, consumers leverage the existing association of a good with the identity of a group.

This finding also reveals that the participants' perception of belonging is not fixed, in that they refer to belonging as a variety of ways of being connected to a group. Interviewees do not always refer to a group as a closed membership, but as an affiliation with relative boundaries, or as the experience of feeling identified or connected with people they also regard as identified with the group. It is from this fluid understanding of group affiliation that, sometimes, the consumer can self-confirm as a member of a group by using a product, with no need of confirmatory cues from others.

**Discussion**
Hammersley (2002) argued that a theory must apply to all circumstances where the specified conditions are met. This does not seem to be the case with the BP theory proposed by Arias and Otnes (2016, 2017) after being empirically tested for the presence of its relevant theoretical conditions, along with conditions noted as not being met. The results of this study suggest that BP is transferable but with limitations, as it describes some aspects of consumption behaviour when seeking to promote belongingness of individuals from a different social and cultural context than that of the original study. Despite transferability, findings reveal the BP model does not account for all the events of belonging-oriented consumption. There are elements that consumers do not take into consideration in their consumption behaviours, others that they recognize only loosely, and others not considered in BP. In particular, the recognition of belongingness that is not conditioned on the confirmation by a third party, even if this third party is affiliated with a belonging target group. This study contributes to identifying the constraints of the BP model, and based on it delves deeper into the phenomenon, recognizing additional subtle events in the behaviour of individuals and accordingly proposes an alternative model, termed the Belonging-Oriented Consumption Model.

This model focuses on the consumer's perspective about their consumption and belongingness decisions consciously aimed to support group belonging, the definition of the group with which they wish to affiliate and the use of brands or products that are functional to that goal and available to them. The elements of the model and its nomenclature are consistent with this approach. The proposed model is depicted in Figure 1:

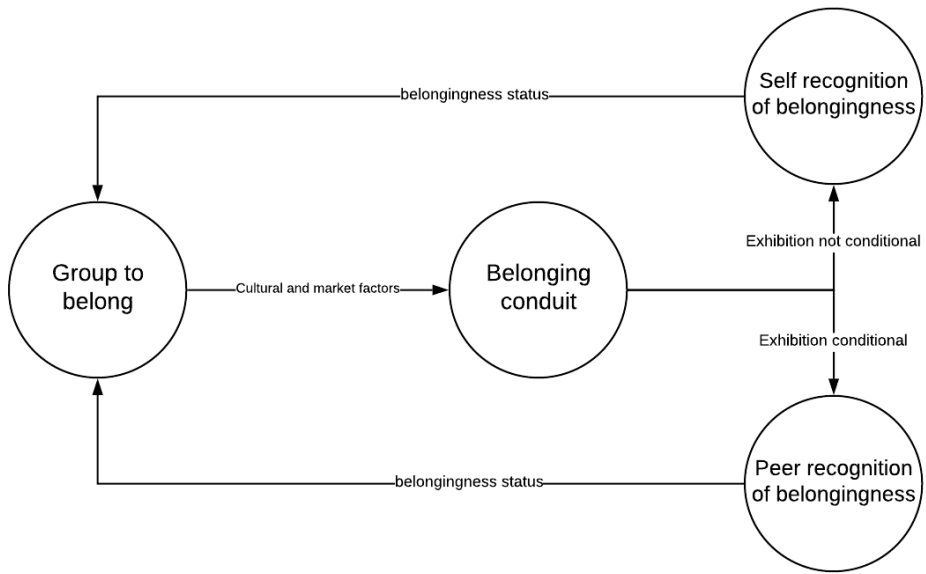

Figure 1. Belonging-Oriented Consumption Model

The model is structured upon the designation of the group to which a person wishes to belong (Group to belong) and on the Belonging conduit, as an instrument that the person uses to link him/herself to the group to which he/she aspires to belong, or to signify that he/she is a member of that group. The selection of a conduit is accordant with the goods or brands associated with or recognised by individuals affiliated to the Group to belong. And this selection is conditioned by cultural factors (cognitions and behavioural traits related to their values and cultural norms) shared by these affiliates and the individual seeking membership status, as well as by marketplace factors because the consumption of products and their use as conduits can only take place among those goods that are accessible in the market for all the actors involved. These remarks are consistent with findings from Vargas-Bianchi (2021a, 2021b), who examined the relationship between the use of the social network Instagram, the selection of products and brands to be used as conduits, and how the features of this social network enhance the display of these conduits. The results of both studies show that the shared meanings attributed to brands and products among people favour their choice as conduits, and that these conduits can be formed with any resources available to the consumer.

Although people recognise the groups to which they belong, or wish to belong, these goals do not appear to result from a systematic effort to identify and assess their progress, but rather to be roughly defined process. Individuals follow paths that require little calculation or excessive elaboration. The participants' statements suggest people achieve an experienced sense of belonging because of their proactive use of consumption and goods as a means, rather than a dependence on the precisely planned confirmation they may receive from others.

Once the individual obtains and has control over the Belonging Conduit, he/she discerns whether it is necessary for the conduit to be exhibited, so that its possession or consumption is recognised by other people (Peer recognition of belongingness) or not (Self recognition of belongingness). When it is the case that peer recognition is sought, it is not required that the other persons are, or are not, members of the group to which affiliation is desired. They may be individuals who, on seeing the conduit, may label the individual as being associated with a group. Alternatively, the individual seeking affiliation may choose not to display the conduit, thus not needing the recognition of third parties to validate his or her affiliation status (Self recognition of belongingness). In that case, his or her own perception would suffice to self-categorise him or herself as belonging to a group when in possession of a product, or when performing specific acts of consumption. This finding has also been reported in studies by Vargas-Bianchi (2021a, 2021b). In both cases (Self recognition and Peer-recognition of belongingness), the belongingness status is checked against the brands, products and consumption practices related with those of the members of the Group to belong. The two conditions are not mutually exclusive, since if someone confirms group affiliation that would be a positive sign of their status, even though the individual concerned may not require such confirmation. These observations are consistent with those of Heatherton (2011) who stated that if human beings need to belong, then they must also have a mechanism aimed at detecting the state of inclusion. Such a mechanism would allow an individual to distinguish the belonging status of a third party, even if this individual is not a group member.

The proposed model of belonging consumption behaviour is consonant with the findings of previous studies on the influence of interpersonal relationships and the shaping of the individual's self and consumption behaviour (Griskevicius, Goldstein, et al. 2006; Park & Feinberg, 2010; Belk, 2017). Such as group conformity (Beran, et al., 2014; Grotts & Johnson, 2013), adherence to group behaviour (Bettman & Escalas, 2005; Algesheimer, Dholakia, & Herrmann, 2005), and the need to belong (Baumeister & Leary, 1995; Nadkarni & Hofmann, 2012), all described in the literature review. The proposed model builds on this knowledge and on the findings from reviewing BP, and distinguishes itself by outlining the belonging consumption process, and its components, by which individuals use consumption to promote group affiliation.

Concerning managerial implications about this study and the model, advertisers aim to create specific meanings for their brands, but consumers are active audiences and the use and meaning they attach to products and brands can be varied. Thus, managers can pay special attention to uses other than the purposes for which a product was created, which arise from the relationship of individuals with the product. These can influence the personality and distinctiveness of a brand. It seems pertinent for managers and advertisers to understand the more distinctive experiences their consumers have with their brands.

This study was limited to analysing those events where the intention to consume to promote group affiliation is conscious. However, much of our consumption happens in a nuanced and unreflective way, and the same must go for consumption related to belonging. It will be appropriate to continue researching this issue when the consumption decision is one of low involvement or when the purchase decision is preceded by little reflection. Future research may examine the process followed when belonging oriented consumption occurs in a pre-attentive state.

It is recommended that the newly proposed model and constructs be examined individually and in more depth in a future study, to comprehend their characteristics and relationships regarding consumer behaviour. The possibility of analysing whether this theoretical framework describes the purchase to belonging behaviour in diverse age groups remains open. It is also recommended that a future study analyse whether the expected belonging experience happens uniquely or if it accepts diverse degrees of group affiliation.

The participants of this study mentioned specific popular brands used as Belonging conduits, particularly consumer technology products and clothing. As some brands position their products on features aligned with the lifestyle of their consumers (Chernev, Hamilton, & Gal, 2011), using brands to favour belonging is related to preceding research that shows that individuals use products to communicate information about themselves to others (Hammerl et al., 2016; Mead et al., 2011; Ratner et al., 1999; Berger & Heath, 2007; Belk, 1988; Holt, 1997). Further research can continue to examine whether other categories of consumer goods and actions are also preferred when they are instrumentalised to favour group affiliation.